\documentclass[fleqn,10pt]{wlscirep}
\usepackage{graphicx} 
\title{L\'evy random walks on multiplex networks}

\author[1,2,3]{Quantong Guo}
\author[3]{Emanuele Cozzo}
\author[1,2,4]{Zhiming Zheng}
\author[3,5,6,*]{Yamir Moreno}

\affil[1]{School of Mathematics and Systems Science, Beihang University, Beijing 100191, China}
\affil[2]{Key Laboratory of Mathematics Informatics Behavioral Semantics(LMIB), Ministry of Education, China}
\affil[3]{Institute for Biocomputation and Physics of Complex Systems (BIFI), University of Zaragoza, Zaragoza 50018, Spain}
\affil[4]{School of Mathematical Sciences, Peking University, Beijing 100191, China}
\affil[5]{Department of Theoretical Physics, University of Zaragoza, Zaragoza 50009, Spain}
\affil[6]{Complex Networks and Systems Lagrange Lab, Institute for Scientific Interchange, Turin, Italy}

\affil[*]{yamir.moreno@gmail.com}

\begin{abstract}
Random walks constitute a fundamental mechanism for many dynamics taking place on complex networks. Besides, as a more realistic description of our society, multiplex networks have been receiving a growing interest, as well as the dynamical processes that occur on top of them. Here, inspired by one specific model of random walks that seems to be ubiquitous across many scientific fields, the L\'evy flight, we study a new navigation strategy on top of multiplex networks. Capitalizing on spectral graph and stochastic matrix theories, we derive analytical expressions for the mean first passage time and the average time to reach a node on these networks. Moreover, we also explore the efficiency of L\'evy random walks, which we found to be very different as compared to the single layer scenario, accounting for the structure and dynamics inherent to the multiplex network. Finally, by comparing with some other important random walk processes defined on multiplex networks, we find that in some region of the parameters, a L\'evy random walk is the most efficient strategy. Our results give us a deeper understanding of L\'evy random walks and show the importance of considering the topological structure of multiplex networks when trying to find efficient navigation strategies.
\end{abstract}
\begin{document}

\flushbottom
\maketitle
%
%
\thispagestyle{empty}


\section*{Introduction}
The study of networks has experienced a burst of activity in the last two decades\cite{Network2, Network, Network3, Network4}. Many diverse dynamical processes have been explored on top of networks, including diffusion processes\cite{Diffusion, cascade, diffusion2}, synchronization\cite{Synchronization, synchronization2}, percolation\cite{percolation, percolation2}, to cite just a few\cite{Critical}. Among these dynamical processes, owning to their wide applications in many scientific fields, including financial time series analysis\cite{financial}, social sciences\cite{social}, genetics\cite{biological} among others, random walks have been attracting more and more attention\cite{Randomwalk, Weighted, Stoequation, Timevarying, Temporal, randomwalk}. Random walks can be used to study transport and to develop different sorts of searching algorithms on networks, with the aim of finding optimal navigation strategies\cite{OptimalDesign, OptimalDesign2, Navigation}. A diversity of random walk processes can be defined and studied, however, most of them rely on the classical random walk process whose dynamics occurs according to the topology of the network\cite{Randomwalk}. In the later scenario, the random walker can only hop to one of the nearest neighbours of the node where it is at any given time, with some -generally the same- probability. Another common random walk process, named L\'evy flight, represents the best strategy for randomly searching a target in an unknown environment. This latter kind of random walk dynamics has been widely observed in many animal species\cite{Levy, LevyHuman}. In its simplest schematization, this stochastic process could drive a walker over very long distances in a single step event that is called `flight'\cite{LevyReview}. The length of the jump, $l$, obeys a power-law probability distribution in the form of $P(l)\sim l^{-\alpha}$\cite{Levy}, which makes it possible for the random walker to hop from one node to any other node.

On the other hand, multiplex networks\cite{multiplex, multiplex2, multiplex3}, i.e., networks composed by many different layers of interactions, are gaining much attention recently. The social and technological revolution brought by the Internet and mobile connections, chats, on-line social networks, and a plethora of other human-to-human machine mediated channels of communications have revealed the need to consider that networks might be made up by many different layers of interactions. The same occurs in other fields, like in contemporary biology, where the needs to integrate multiple sets of omic data naturally leads to a multiplex network as a schematization of the system under study. Also in the traditional field of transportation networks, the concept of multiplex networks has a natural translation in different modes of transportations connecting the same physical location in a city, a country, or on the globe. Finally, in the area of engineering and critical infrastructures, it applies to the interdependence of different lifelines \cite{transport}. Furthermore, research shows that the topological and dynamical properties of a multiplex network are in general different as compared to those of a single layer network \cite{multitopo,sanchez2014dimensionality, multitopo3, multitopo4}, as well as the dynamical processes on it\cite{Cozzo2013,multidyna2, multidyna3}. For example, it has been shown that a diffusion process can have an enhanced-diffusive behaviour\cite{diffusion2} on a multiplex network, which means that the time scale associated to it is shorter than that occurring on a single layer network.

\begin{figure}[htbp]
\centering
\includegraphics[width=1.0\columnwidth]{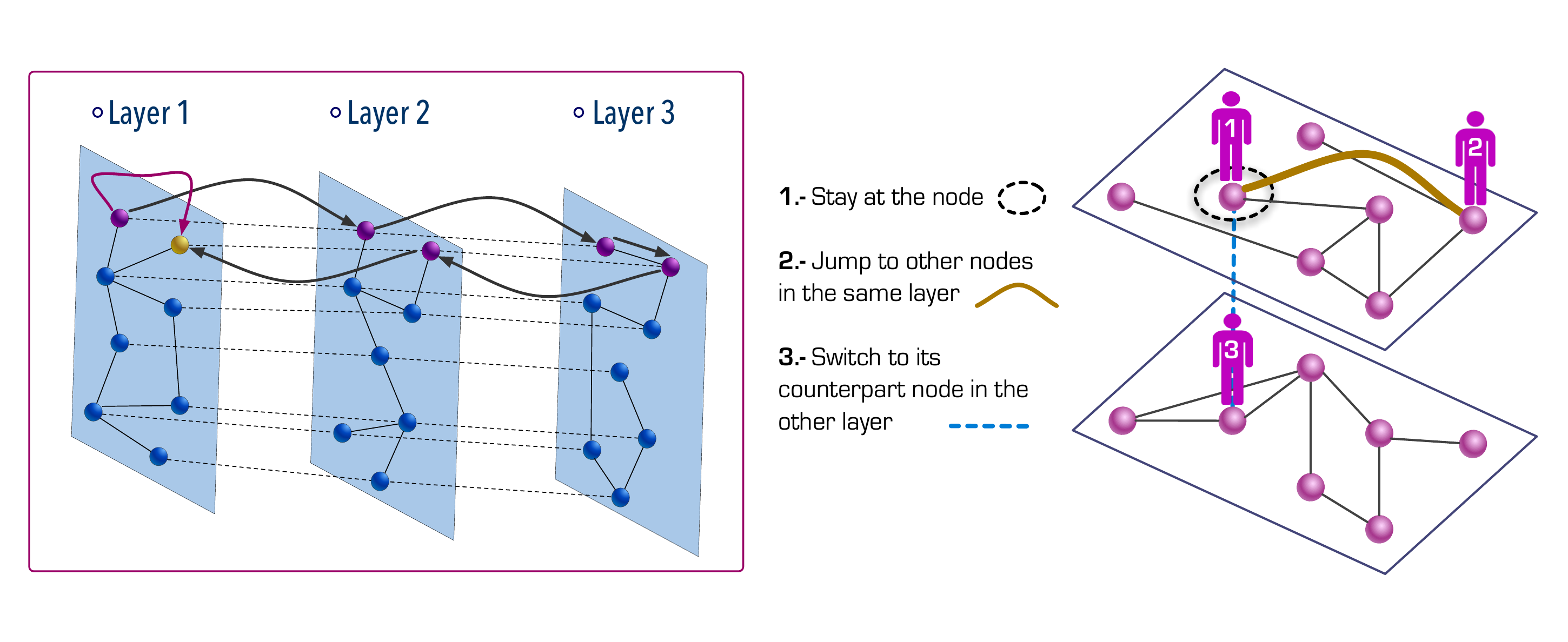}
\caption{Illustration of the L\'evy flight navigation strategy on a multiplex network. In the toy model, we consider a three-layer multiplex network and show two different paths that can lead the walker to the yellow node (one involves a L\'evy flight and the other implies that the walker follows the topological path of the graph). The right panel summarizes the different elementary steps that a walker can adopt in our model as indicated.}
\label{toymodel}
\end{figure}

All the already existing studies of random walks on multiplex networks adopt a nearest-neighbour navigation strategy\cite{DomenPANS, multirandom}. The aim of this paper is to generalize L\'evy flights random walks to multiplex networks, which means that a random walker has a certain probability to move to any other node without the need of a direct connection as far as the network is concerned. At each step, the random walker has three options: the first one is to stay at the same node, the second one is to jump to other nodes on the same layer and the last one is to switch to one of its counterparts on other layers, as illustrated in Fig.\ref{toymodel}. According to the definition of the dynamics, we obtain the expression for the stationary distribution and the random walk centrality\cite{Randomwalk}. Besides, with the help of stochastic matrix theory\cite{Weighted, Graphbook},  we derive the exact expression for the mean first passage time (MFPT). The MFPT is used to describe the expected time needed for a random walker starting from a source point to reach a given target point\cite{MFPT}. Finally, we also compare the results for L\'evy flights with other random walks dynamics obtained also on multiplex networks\cite{DomenPANS}, finding that, under certain conditions, the L\'evy random walk is the most efficient from a global viewpoint.

\section*{Results}
In this work we consider undirected connected node-aligned multiplex networks \cite{multiplex2}. A node-aligned multiplex network is made up of $L$ layers with N nodes $i=\{1,2,\dots,N\}$ on each layer. An adjacency matrix $A_\alpha=\{a^\alpha_{ij}\}_{N\times N}$, with $\alpha=\{1,2,...,L\}$, is associated to each layer $\alpha$. Besides, a coupling matrix $C=\{c^{\alpha\beta}_{ij}\}_{NL\times NL}$ describes the coupling between nodes in different layers; since each node is coupled only with its counterparts in different layers, then, only the elements of the type  $c^{\alpha\beta}_{ii}$ are different from zero.

The whole multiplex network can be described by the supra-adjacency matrix $A=\left\lbrace a_{ij}\right\rbrace _{NL\times NL}=\bigoplus A_\alpha + C$ \cite{multiplex2}. Additionally, we consider another set of matrices associated with the multiplex network, that is, we consider a distance matrix $D_{\alpha}=\{d^{\alpha}\}_{N\times N}$ associated to each layer $\alpha$, where the element $d_{ij}^{\alpha}$ encodes the length (number of steps) of the shortest path connecting node $i$ to node $j$ in layer $\alpha$.\cite{Levy} We indicate the probability to find a random walker in node $j$ of layer $\beta$ at time $t$ starting from node $i$ of layer $\alpha$ at $t=0$ by $p_{ij}^{\alpha\beta}(t)$. The discrete-time master equation is given:

\begin{equation}
p_{ij}^{\alpha\beta}(t+1)=\sum_{m=1}^{N}p_{im}^{\alpha\beta_{t}}(t)w_{mj}^{\beta_{t}\beta},
\end{equation}
where $w_{ij}^{\alpha\beta}$ is the transition probability of moving from node i of layer $\alpha$ to node j of layer $\beta$.\\
To account for the inter-layer connections, we introduce $D_{ii}^{\alpha\beta}$ to quantify the "cost" to switch from layer $\alpha$ to layer $\beta$ at node i, while $D_{ii}^{\alpha\alpha}$ quantifies the "cost" of staying in the same node and in the same layer.

We can now define the transition probabilities $w_{ij}^{\alpha\beta}$ to be  

\begin{equation}
    w_{ij}^{\alpha\beta}=
   \begin{cases}
   \dfrac{(d_{ij}^{\alpha})^{-\theta}}{s_{i}^{\alpha}} &\mbox{ $\alpha=\beta, \left| i-j\right| \% N\neq 0 $}\\
   \\
   \dfrac{D_{ii}^{\alpha\alpha}}{s_{i}^{\alpha}} &\mbox{ $\alpha=\beta, i=j$}\\
   \\
   \dfrac{D_{ii}^{\alpha\beta}}{s_{i}^{\alpha}} &\mbox{ $\alpha \neq \beta, \left| i-j\right| \% N =0$}
   \end{cases}
\end{equation}
where $s_{i}^{\alpha}=\sum_{j}(d_{ij}^{\alpha})^{-\theta}+\sum_{\beta}D_{ii}^{\alpha\beta}$ is the strength of node $i$ with respect to its connections in the multiplex network, which takes into account the probability of staying at this node and of switching to another layer. As in the case of traditional single layer networks \cite{Levy}, the transition probabilities $w_{ij}^{\alpha\beta}$ define a dynamical process where a random walker can visit not only the nearest neighbours of a node, but also nodes without direct connections with it on the same layer, while the random walker can switch layer only staying at the same node. Since $(d_{ij}^{\alpha})^{-\theta}$ has an exponential decay according to the shortest path between the source node and the target node, the farther they are, the smaller the probability to hop to one from the other. The parameter $\theta$, which varies in the range $[0, \infty)$, controls the decay of this probability.

In the following we will derive the mean first passage time (MFPT), that is a characteristic quantity related to a random walk\cite{Weighted}. By iterating Eq.(1),we get an explicit expression for $p_{ij}^{\alpha\beta}(t)$: 
\begin{equation}
p_{ij}^{\alpha\beta}(t)=\sum_{j_{1},...j_{t-1}}w_{ij_{1}}^{\alpha\beta_{1}}...w_{j_{t-1}j}^{\beta_{t-1}\beta}
.
\end{equation}
Comparing $p_{ij}^{\alpha\beta}(t)$ with $p_{ji}^{\beta\alpha}(t)$ according to the definition in Eq.(2), we get
\begin{equation}
p_{ij}^{\alpha\beta}(t)s_{i}^{\alpha}=p_{ji}^{\beta\alpha}(t)s_{j}^{\beta}
\end{equation}
For the stationary solution, which corresponds to the infinite time limit, we can get $lim_{t\rightarrow\infty}p_{ij}^{\alpha\beta}(t)=p_{j}^{\beta}$\cite{Levy}. Hence, Eq.(4) implies that $p_{j}^{\beta}s_{i}^{\alpha}=p_{i}^{\alpha}s_{j}^{\beta}$ and the probability $p_{i}^{\alpha}$ reduces to
\begin{equation}
p_{i}^{\alpha}=\dfrac{s_{i}^{\alpha}}{s}
\end{equation}
where $s=\sum_{\alpha}\sum_{i}s_{i}^{\alpha}$ characterises the strength of the whole multiplex network. The expression of the stationary distribution $p_{i}^{\alpha}$ shows that the larger the strength of node $i$, the more often it will be visited, which is valid for any undirected network\cite{Book}.

The average of the MFPT over the stationary distribution is (see Methods for details of the derivation)
\begin{equation}
\langle T\rangle=\sum_{k=2}^{NL}\dfrac{1}{1-\lambda_{k}},
\end{equation}
whew $\lambda_k$ are the eigenvalues of the matrix  $\mathrm W=\left\lbrace w_{ij}^{\alpha\beta}\right\rbrace_{NL\times NL}$, with $1=\lambda_{1}>\lambda_{2}\geqslant\cdots \geqslant\lambda_{NL}\geqslant-1$.
Besides, the random walk centrality of node $i$, as introduced in \cite{Randomwalk}, is $C_{i}^{\alpha}=(\tau_{i}^{\alpha})^{-1}$, where $\tau_{i}^{\alpha}$ is defined as $\displaystyle\sum_{t=0}^{\infty}\{p_{ii}^{\alpha\alpha}(t)-p_i^\alpha\}/p_{i}^{\alpha}$. $\tau_{i}^{\alpha}$ is given by (see Methods)
\begin{equation}
\tau_{i}^{\alpha}=\sum_{k=2}^{NL}\dfrac{1}{1-\lambda_{k}}\dfrac{\phi_{ki}^{2}}{\phi_{1i}^{2}}
\end{equation}
Hence, we have derived the exact expression of the transition probability $p_{ij}^{\alpha\beta}$ and the MFPT $\langle T\rangle$.  In addition, in order to analyse the navigation of L\'evy walks, we average all the $\tau_{i}^{\alpha}$'s over the whole network, which means $\tau=\dfrac{1}{NL}\sum_{i,\alpha}\tau_{i}^{\alpha}$. Note that with respect to the local index $\tau_{i}$, which represents the average time needed to reach node $i$ from a randomly chosen node, $\tau$ gives the average number of steps needed to reach any node independently of the initial condition\cite{Levy}.

\begin{figure}[htbp]
\centering
\includegraphics[width=0.5\columnwidth]{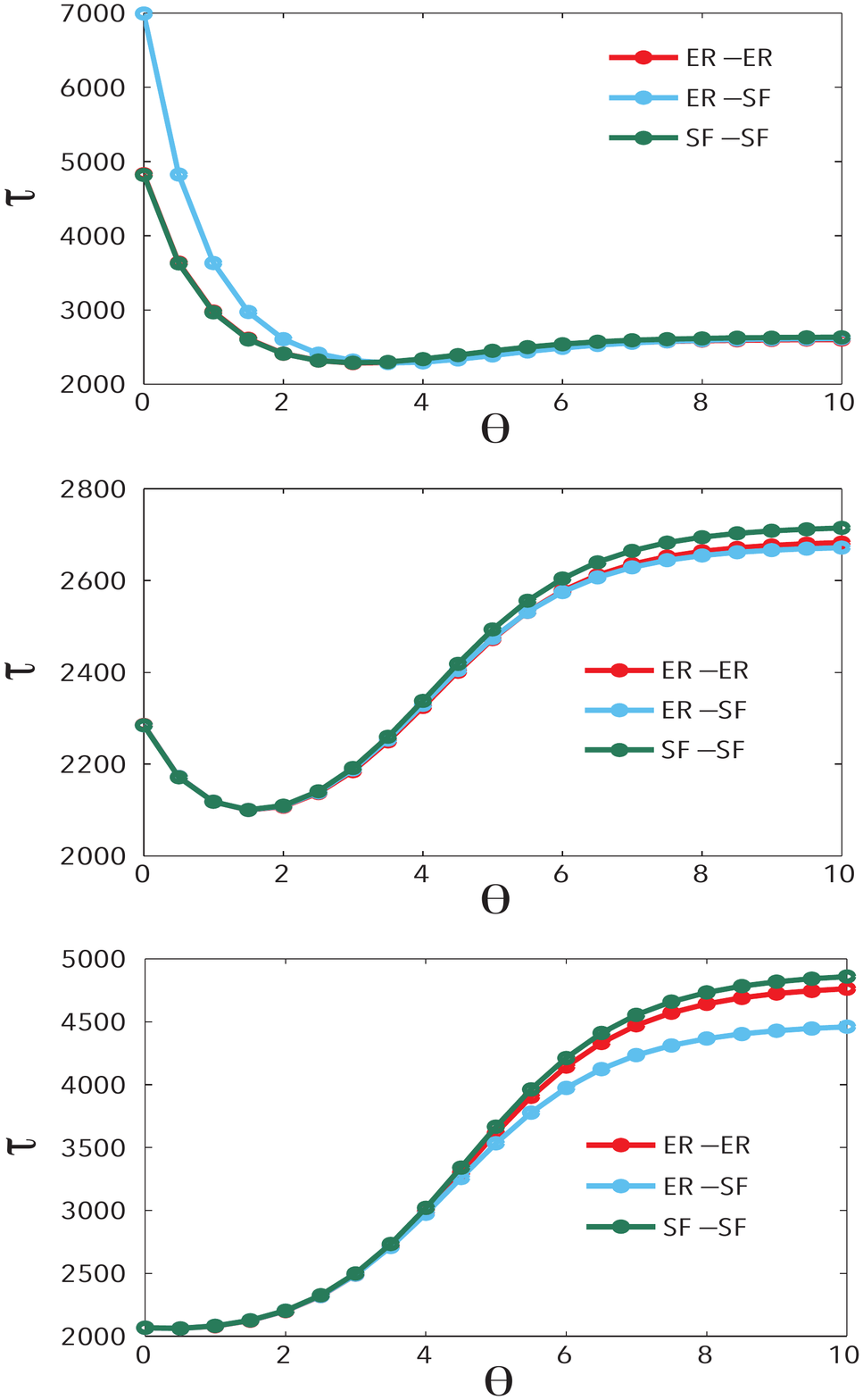}
\caption{The quantity $\tau$ vs the L\'evy flight index $\theta$ for different two-layer multiplex networks. Each layer is an ER network or a SF network with 1000 nodes as indicated in the legends. The values of the coupling strength $D_{X}$ between the two layers are: (a) $D_{X}$=0.1, (b) $D_{X}$=1, (c) $D_{X}$=10.}
\label{fig:Rvalue}
\end{figure}

Next, we proceed to characterise L\'evy random walks on multiplex networks drawing on the exact analytic results given above. For the sake of simplicity, following Ref\cite{DomenPANS}, we assign the same value $D_{X}$ to all the $D_{ii}^{\alpha\beta}$, i.e., switching layers has the same cost at any node. In Fig.\ref{fig:Rvalue} we show  $\tau$ vs $\theta$ for different topologies and different values of the cost $D_{X}$. It is worth noting that, while for large values of the parameter $D_{X}$ the behaviour of the time $\tau$ when varying $\theta$ is qualitatively similar to the classical case of single layer networks, for small  and intermediate values of $D_{X}$ it deviates significantly from the classical case. In particular,  the relationship between $\tau$ and $\theta$ appears to be of three different kinds depending on $D_{X}$: when $D_{X}$ is sufficiently small ($D_{X}=0.1$ in panel (a)) $\tau$ decreases quickly for small value of $\theta$, while it remains more or less constant for large values of $\theta$, ; when $D_{X}$ is sufficiently large ($D_{X}=10$ in panel (c)) $\tau$ increases monotonically with $\theta$, as in classical single layer networks, with the speed of the increase being much smaller when $\theta$ is small. Furthermore, for intermediate values of $D_{X}$ ( $D_{X}=1$ in panel (b)) $\tau$ shows a clear minimum for a given value of $\theta$. This phenomenology, that is, the fact that the efficiency depends on the coupling, constitutes the central finding of our study. 

In single layer networks, setting $\theta=0$ is always the best strategy to navigate a network, as the global time $\tau$ is minimum. However, in multiplex networks the value of $\theta$ -it's optimal- that minimizes $\tau$ depends on the value of the coupling $D_{X}$. Interestingly enough, for low values of $D_{X}$, the limiting case $\theta\rightarrow\infty$, which corresponds to the normal random walks on networks, can be more efficient than $\theta=0$. Other scenarios worth inspecting are given by the topologies of the networks that made up each layer of the multiplex. In particular, a multiplex network can be made up of different combinations of homogeneous (Erdos-Renyi (ER)) and heterogeneous (Scale-Free (SF)) networks. We have also explored these scenarios numerically for different regimes of the coupling parameter. For $D_{X}\ll=1$, different structures lead to different relationships between $\tau$ and $\theta$. When $\theta$ is small, a SF-SF multiplex network has a much smaller $\tau$ than an ER-ER or an ER-SF multiplex network; however, if $D_{X}$ is bigger than 1, the difference is evident only when $\theta\gg 0$. Altogether, the previous results show that whether the optimal value of the L\'evy walk index $\theta$ is constant across different multiplex topologies depends on the value of the coupling strength $D_{X}$.

\begin{figure}[htbp]
\centering
\includegraphics[width=0.8\columnwidth]{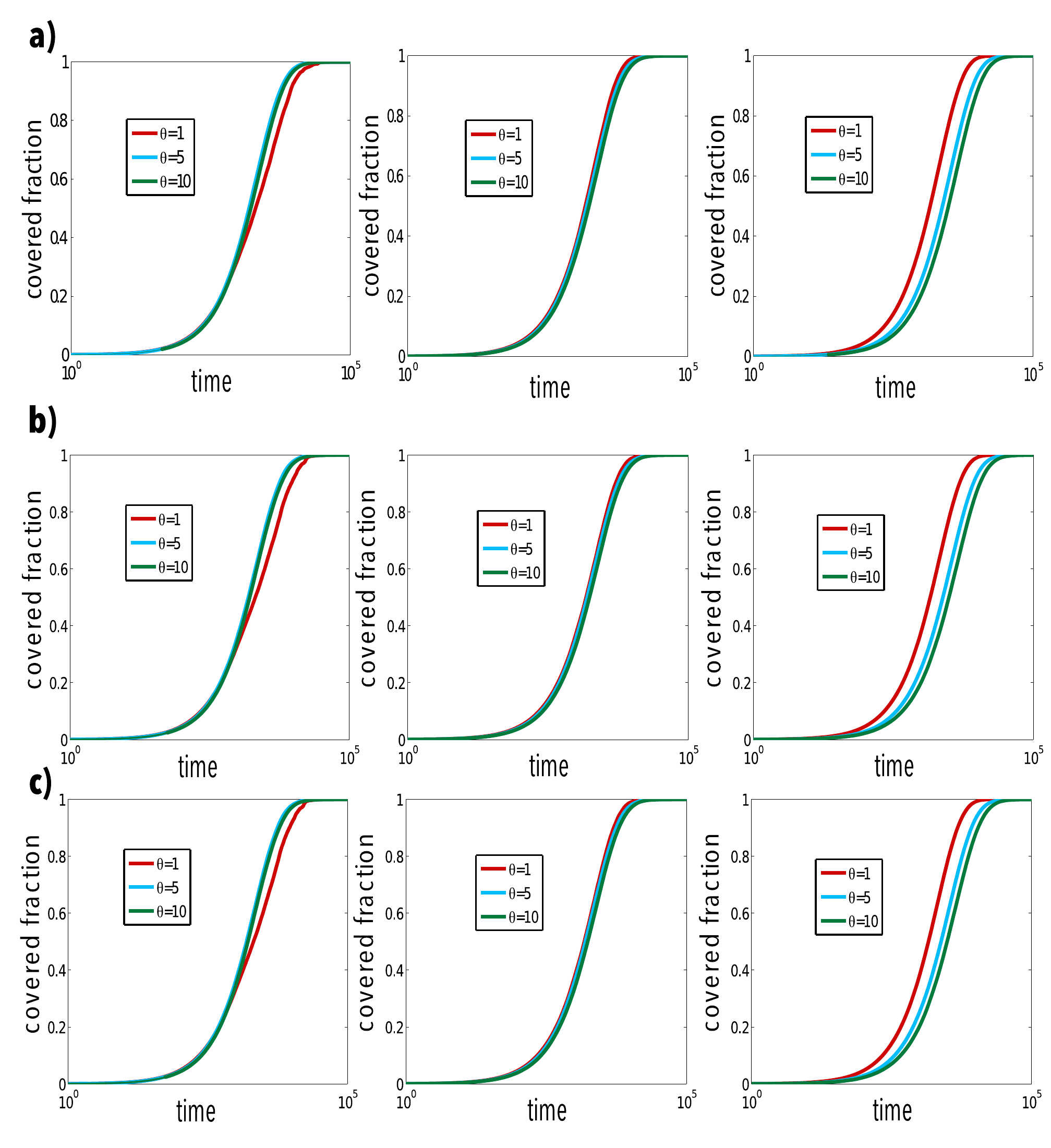}
\caption{Number of visited nodes versus time for L\'evy random walks on multiplex networks. The structures of the multiplex networks considered are: ER-ER (top panels a), ER-SF (middle panels b) and SF-SF (bottom panels c). In each configuration, the synthetic networks of each layer are composed of $10^3$ nodes. From left to right, the values of $D_X$ are 0.1, 1, 10. The L\'evy index $\theta$ used are indicated in the legend.}
\label{covered}
\end{figure}

Next, in order to provide more numerical evidences of what we have found analytically, we present the results obtained when the fraction of covered nodes is taken into consideration. Figure \ref{covered} shows this magnitude as a function of time for different multiplex networks. In the first case (panel a), the network is made up of two ER networks with the same structure. For a small value of $D_{X}$ (upmost left figure), it can be seen that the bigger the value of $\theta$ is, the higher the efficiency of the L\'evy random walk is. This also confirms the results in Fig.\ref{fig:Rvalue}, since a larger value of $\theta$ leads to a smaller value of $\tau$. With respect to other values of $D_{X}$, such as $D_{X}=1$ (middle figures in all panels), $D_{X}=10$ (upmost right figures in all panels), the results for $\tau$ in Fig.\ref{fig:Rvalue} are also confirmed. In the case of other kinds of arrangements for the networks at each layer, we obtain the same results, as can be seen from panels b and c in Fig.\ref{covered}. Furthermore, the comparison of the results obtained for different combinations shows that the topologies of the networks in each layer do not play a significant role.

\begin{figure}
\centering
\includegraphics[width=1.0\columnwidth]{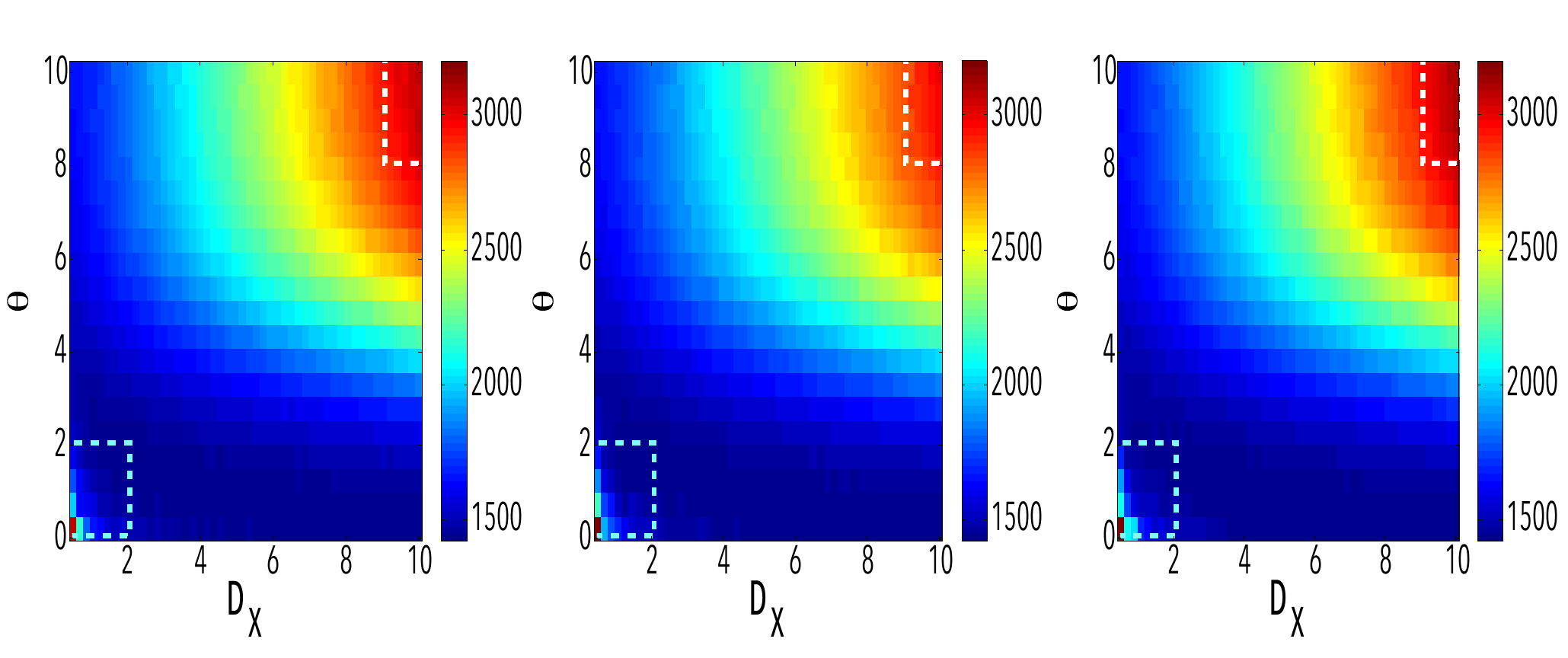}
\caption{The effects of inter-layer weight $D_X$ and L\'evy index $\theta$ on the efficiency of L\'evy random walks. The color-coded map describes the time needed to cover 50\% of all the nodes ($10^3$) in each layer. From left to right, the structure of network in each layer is ER-ER, ER-SF and SF-SF, respectively. The rectangles highlights the areas that show the largest differences due to the multiplex structure of the system.}
\label{heatmap}
\end{figure}

The previous results indicate that the coupling strength between layers is a crucial factor determining the structure and the dynamical behavior of the system\cite{Abrupt}.  In addition, as described above, being used to characterize the cost for a walker to switch between layers, the value of $D_{X}$ also has distinct effects on L\'evy random walks on multiplex networks. In order to further explore the details of these effects, as a function of $\theta$, we show in Fig.\ref{heatmap} the time needed to cover the 50\% of all nodes as a function of both$D_{X}$ and $\theta$. As shown in the figure, an interesting phenomenology appears. Firstly, the highest values of the time needed to cover half of the network locate at the up-right and down-left corners, where the values of $D_{X}$ and $\theta$ are the biggest and the smallest. Moreover, in a significant range of values of $\theta$, increasing $D_{X}$ does not change greatly the time needed to cover 50\% of the nodes. However, if $\theta$ is large enough, the increase of $D_{X}$ have a larger impact. Note that these results also confirm the analytical findings about $\tau$, as can be seen in Fig.\ref{fig:Rvalue}.

\begin{figure}[htbp]
\centering
\includegraphics[width=1.0\columnwidth]{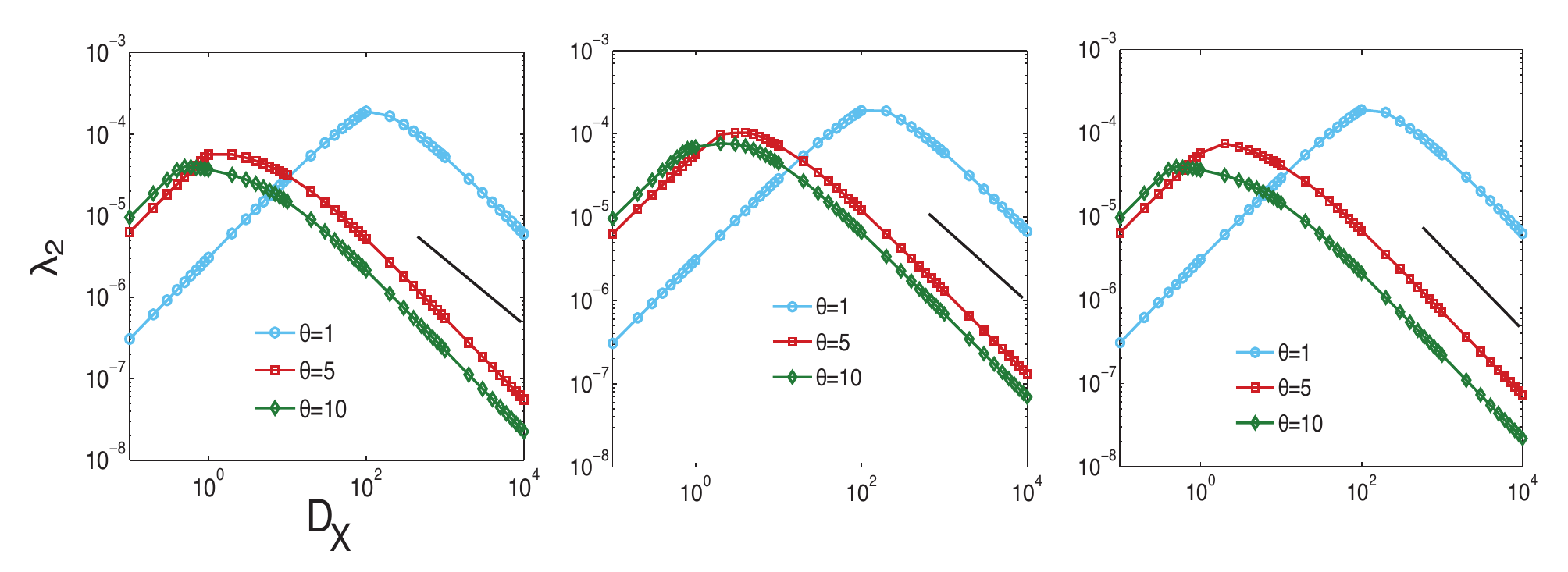}
\caption{The second smallest eigenvalue of the generalized supra-laplacian matrix as a function of the inter-layer weight $D_X$ for three different multiplex topologies, which from left to right are ER-ER, ER-SF, SF-SF, respectively. Each panel describes L\'evy random walk with different L\'evy index $\theta$ (1, 5, 10). The solid line corresponds to $D_{X}^{-1}$.}
\label{eigenvalue}
\end{figure}

Another result worth highlighting that connects our results for $\tau$ with the structural properties of the multiplex involves the second largest eigenvalue $\lambda_2$. As shown for the smallest eigenvalue of the supra-Laplacian \cite{Abrupt}, there is a transition point that separates two different regimes in interdependent networks: in one regime, all the layers are structurally decoupled and in the other regime, the system behaves as a single layer. The same result holds for the second smallest eigenvalue of the generalized supra-Laplacian of a multiplex network when increasing the coupling strength $D_{X}$. Specifically, the generalized supra-Laplacian is

$A=
\begin{pmatrix}
D^{11}I+W^{(1)} & D^{12}I & \cdots & D^{1L}I\\
D^{21}I & D^{22}I+W^{(2)} & \cdots & D^{2L}I\\
\vdots & \vdots & \ddots & \vdots\\
D^{L1}I & D^{L2}I & \cdots & D^{LL}I+W^{(L)}
\end{pmatrix}$
with $W^{(\alpha)}=\left\lbrace {w_{ij}^{\alpha \alpha}}\right\rbrace _{L\times L}$ and $I$ is the $N\times N$ identity matrix.

Figure\ref{eigenvalue} shows the dependency of $\lambda_2$ with $D_{X}$ for different values of the L\'evy flight parameter $\Theta$. Also in this case (see \cite{DomenPANS}), $\lambda_2\propto D_{X}^{-1}$, regardless of the network structure as showed it can be seen in the different panels of Fig.\ref{eigenvalue}. Finally, for the sake of comparison with results obtained for other random walk dynamics, we compare their efficiency\cite{DomenPANS} with that of the L\'evy flight. In the first case (RWC), the random walker in node $i$ can move to any one of its neighbors $j$ on the same layer with the transition probability $w_{ij}^{\alpha \alpha} =\dfrac{1}{k_{i}}$, where $k_{i}$ is the degree of node $i$. Secondly, we also consider the case (RWD) in which the random walker is allowed to jump to any other node with probability $w_{ij}^{\alpha \alpha}=\frac{s_{i}^{\alpha}}{s_{max}}$, where $s_{max}=\text{max}_{\{i,\alpha\}}s_{i}^{\alpha}$. Lastly, a third scenario (RWP) considers that it is possible for a random walker to switch layers and jump to another neighborhood at the same step.

\begin{figure}
\centering
\includegraphics[width=0.6\columnwidth]{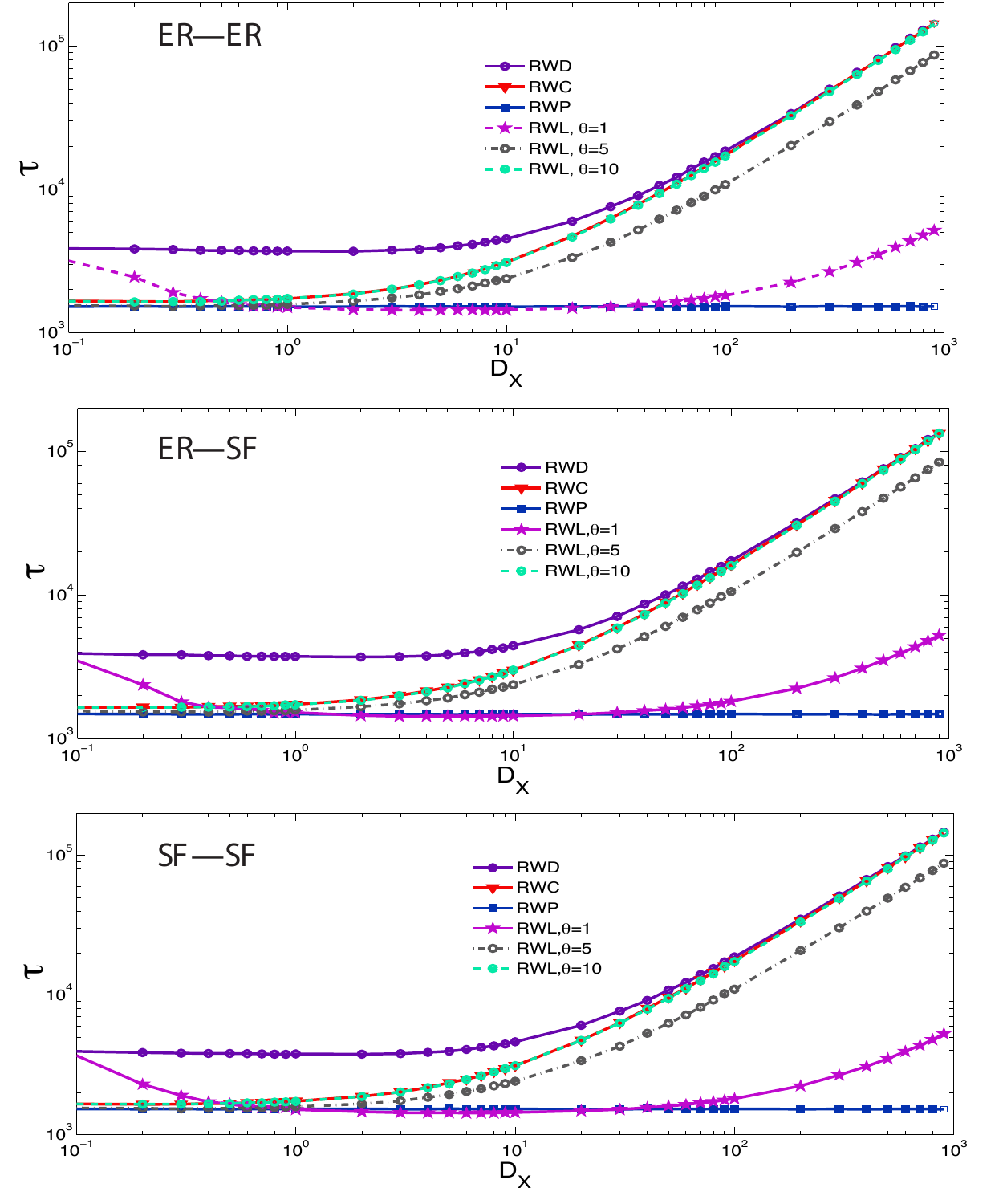}
\caption{Time needed to cover 50\% of all nodes on three types of multiplex networks, as a function of the inter-layer weight $D_{X}$. We compare the results obtained for the L\'evy random walk (RWL) studied here with three other scenarios for the walks, as discussed in the text. The values of $\theta$ considered are 1, 5, 10, respectively. Each layer has $10^3$ nodes and all the simulations were averaged over 100 realizations.}
\label{all}
\end{figure}


In Fig. \ref{all}, we show the time $\tau$ needed to cover 50\% of all the nodes as a function of the value of $D_{X}$ with respect to different topological structures. For the L\'evy random walk, we study three different cases where the index $\theta$ equals 1, 5, 10, respectively.  Comparing the L\'evy case with the three others mentioned above, one can get further insights on the different strategies for navigation, that is to say, there is no strategy that is always the most efficient for any network and an arbitrary coupling strength. For instance, taking the L\'evy random walk as an example: when $\theta$ is small ($\theta=1$), the time $\tau$ appears to be the smallest in the range $1<D_{X}<10$, but as $\theta$ if further increased, the time needed to cover the 50\% of the nodes of the network is almost the same as compared to that needed for a classical random walk, which in its turn is not the most efficient. This is easy to understand because in a L\'evy random walk, when $\theta\rightarrow \infty$, the transition probability $w_{ij}=\dfrac{d_{ij}^{-\theta}}{s_i}\rightarrow 0$ if the shortest path $d_{ij}$ is larger than 1. Therefore, the fist of the cases to which we compare -i.e., the classical random walk- is a special case of a L\'evy walk when $\theta\rightarrow \infty$\cite{Levy}.






\section*{Discussion}

In summary, we have studied L\'evy random walks on multiplex networks. With the help of stochastic matrix theory, we have calculated analytically the expression of the stationary distribution and MFPT from any node to any other node. Besides, we have also obtained an exact expression of the average time $\tau$ needed to reach a node regardless of the source node. This dynamics on multiplex networks shows a strong dependence on the inter-layer weight $D_X$ and the L\'evy index $\theta$. Our main result is that when $D_X$ is small enough, contrary to the case of a L\'evy random walk on single layer networks, the bigger the index $\theta$ is, the more efficient the L\'evy random walk is. In order words, in that region of parameter values, although it is not very likely for any given walker to jump directly to other nodes far away, the total average time $\tau$ needed to visit any node independently of the initial condition is smaller. Interestingly, if the value of $\theta$ is not too large, for instance for $\theta<4$, $D_X$ does not have a significative impact on $\tau$. The present results add to previous works that explored other kinds of random walk processes on multiplex networks, and allow to have a more complete picture that highlights the importance of considering the interconnected nature of many systems if we aim at finding the best navigation strategies and develop searching and navigability algorithms for such interdependent networked systems.

\section*{Methods}

In the following, by using the formalism of generating functions\cite{generatef}, we will get the analytical result for MFPT. The first passage probability $q_{ij}^{\alpha\beta}(t)$ from node i of layer $\alpha$ to node j of layer $\beta$ after $t$ steps satisfies the relation
\begin{equation}
p_{ij}^{\alpha\beta}(t)=\sum_{l=0}^{t}q_{ij}^{\alpha\beta}(l)p_{jj}^{\beta\beta}(t-l)
\label{first_passage_probability}
\end{equation}
Let ${T}_{ij}^{\alpha\beta}$ denote the MFPT from node i of layer $\alpha$ to node j of layer $\beta$, then
\begin{equation}
{T}_{ij}^{\alpha\beta}=\sum_{t=0}^{\infty}tq_{ij}^{\alpha\beta}(t)
\end{equation}
here, as proposed in\cite{Weighted}, we introduce the following generating functions:
\begin{equation}
\tilde{Q}_{ij}^{\alpha\beta}(x)=\sum_{t=0}^{\infty}q_{ij}^{\alpha\beta}(t)x^{t}
\end{equation}
\begin{equation}
\tilde{P}_{ij}^{\alpha\beta}(x)=\sum_{t=0}^{\infty}p_{ij}^{\alpha\beta}(t)x^{t}
\label{gf2}
\end{equation}
where $\vert x\vert<1$, inserting Eq.\ref{first_passage_probability} into \ref{gf2}, we get
\begin{equation}
\tilde{Q}_{ij}^{\alpha\beta}(x)=\dfrac{\tilde{P}_{ij}^{\alpha\beta}(x)}{\tilde{P}_{jj}^{\beta\beta}(x)}
\end{equation}
Since ${T}_{ij}^{\alpha\beta}=\displaystyle\sum_{t=0}^{\infty}tq_{ij}^{\alpha\beta}(t)={\dfrac{\mathrm d}{\mathrm dx}\tilde{Q}_{ij}^{\alpha\beta}(x)}\bigg \vert_{x=1}$, the problem of solving for the MFPT is reduced to calculate the derivative of $\tilde{Q}_{ij}^{\alpha\beta}(x)$ and evaluate it at $x=1$.

We will address this point making use of the stochastic matrix theory \cite{Graphbook}. For the sake of simplicity, we use the matrix $\mathrm W=\left\lbrace w_{ij}^{\alpha\beta}\right\rbrace_{NL\times NL}$ and the matrix $\mathrm S=\mathrm {diag}[s_{1}^{1},s_{2}^{1}\cdots s_{N}^{1}\cdots s_{i}^{\alpha}\cdots s_{1}^{L}\cdots s_{N}^{L}]$ to describe the transition probabilities and node strengths, respectively. It is clear that the matrix $W$ is a stochastic matrix, since for any node $i$ its elements satisfy that $\sum_{j=1}^{NL}w_{ij}=1$. $W$ is an antisymmetric matrix. Because of that, we introduce the matrix
\begin{equation}
\Gamma=S^{\frac{1}{2}}WS^{-\frac{1}{2}}=S^{-\frac{1}{2}}(SW)S^{-\frac{1}{2}}
\label{Gamma}
\end{equation}
which is symmetric and similar to W. Thus, they have the same eigenvalues. Since $\Gamma$ can be diagonalized and the eigenvalues are all real, we define $\lambda_{1},\lambda_{2},\cdots \lambda_{NL}$ as its eigenvalues. These eigenvalues satisfy that $1=\lambda_{1}>\lambda_{2}\geqslant\cdots \geqslant\lambda_{NL}\geqslant-1$. Let $\mathrm{\Phi}=\{\phi_{1},\phi_{2},\cdots \phi_{NL}\}$ denote the corresponding normalized, real-valued, and mutually orthogonal eigenvectors. As a result, the matrix $\Gamma$ can be written as
\begin{equation}
\Gamma=\Phi \mathrm{diag}[\lambda_{1},\lambda_{2},\cdots \lambda_{NL}]\Phi^{T}
\end{equation}
which, together with \ref{Gamma}, leads to
\begin{equation}
W=S^{-\frac{1}{2}}\Gamma S^{\frac{1}{2}}=S^{-\frac{1}{2}}\Phi \mathrm{diag}[\lambda_{1},\lambda_{2},\cdots \lambda_{NL}]\Phi^{T}S^{\frac{1}{2}}
\label{W}
\end{equation}
Then considering the master equation Eq.(1), we can get $P(t)=P(0)W^t=S^{-\frac{1}{2}}\Gamma^{t}S^{\frac{1}{2}}$, whose element denoted by $p_{ij}^{\alpha\beta}(t)$ represents the transition probability from node i of layer $\alpha$ to node j of layer $\beta$ in t steps. Note that the elements of the matrix $P(0)$ fulfill the following relations
\begin{equation}
p_{ij}^{\alpha\beta}(0)=
\begin{cases}
1, \qquad \alpha=\beta, i=j\\
0, \qquad \text{else}
\end{cases}
\end{equation}

Then, inserting Eq.\ref{W} into the expression of $P(t)$, one has
\begin{equation}
p_{ij}^{\alpha\beta}(t)=\sum_{k=1}^{NL}\lambda_{k}^{t}\phi_{ki}\phi_{kj}\sqrt{\dfrac{s_{j}^{\beta}}{s_{i}^{\alpha}}}
\label{poft}
\end{equation}
where $\phi_{i}=(\phi_{i1},\phi_{i2},\cdots \phi_{iNL})^{T}$ and they satisfy
$\phi_{i}^{T}\phi_{j}=1$ if $i=j$, else $\phi_{i}\phi_{j}=0$, which means
\begin{equation}
\sum_{k=1}^{NL}\phi_{ik}\phi_{jk}=\sum_{k=1}^{NL}\phi_{ki}\phi_{kj}=0
\label{ref}
\end{equation}

We have now an expression for $p_{ij}^{\alpha\beta}(t)$, plugging it into Eq. \ref{gf2}, it is easy to obtain 
\begin{equation}
\tilde{P}_{ij}^{\alpha\beta}(x)=\dfrac{s_{j}^{\beta}}{s}\dfrac{1}{1-x}+\sum_{k=2}^{NL}\dfrac{1}{1-\lambda_{k}x}\phi_{ki}\phi_{kj}\sqrt{\dfrac{s_{j}^{\beta}}{s_{i}^{\alpha}}},\qquad
\tilde{P}_{jj}^{\beta\beta}(x)=\dfrac{s_{j}^{\beta}}{s}\dfrac{1}{1-x}+\sum_{k=2}^{NL}\dfrac{1}{1-\lambda_{k}x}\phi_{kj}^{2}
\end{equation}

According to the definition given above, the MFPT $T_{ij}^{\alpha\beta}$ can be calculated by differentiating $\tilde{Q}_{ij}^{\alpha\beta}(x)$
\begin{equation}
{T}_{ij}^{\alpha\beta}=\dfrac{s}{s_{j}^{\beta}}\sum_{k=2}^{NL}\dfrac{1}{1-\lambda_{k}}\left( \phi_{kj}^{2}-\phi_{ki}\phi_{kj}\sqrt{\dfrac{s_{j}^{\beta}}{s_{i}^{\alpha}}}\right)
\label{mfptn}
\end{equation}

In addition, using Eq.\ref{mfptn}, we have
\begin{equation}
\langle T\rangle=\sum_{\substack{j=1\\ i\neq j}}^{NL}T_{ij}^{\alpha\beta}p_{j}^{\beta}=\sum_{\substack{j=1\\ i\neq j}}^{NL}\sum_{k=2}^{NL}\dfrac{1}{1-\lambda_{k}}\left( \phi_{kj}^{2}-\phi_{ki}\phi_{kj}\sqrt{\dfrac{s_{j}^{\beta}}{s_{i}^{\alpha}}}\right)
\end{equation}

Using Eq.\ref{ref} and the relation $\phi_{1i}\phi_{1j}\sqrt{\dfrac{s_{j}^{\beta}}{s_{i}^{\alpha}}}=\dfrac{s_{j}^{\beta}}{s}$, which means $\phi_{1j}=\sqrt{\dfrac{s_{j}^{\beta}}{s}}$, we can get 
\begin{equation}
\langle T\rangle=\sum_{k=2}^{NL}\dfrac{1}{1-\lambda_{k}},
\end{equation}

where the time $\langle T\rangle$ is the average of the MFPT over the stationary distribution, obviously it does not depend on i, and it is known as the Kemeny's constant\cite{Levy}.
Besides, as introduced in \cite{Randomwalk}, we can calculate the quantity $C_{i}^{\alpha}=(\tau_{i}^{\alpha})^{-1}$, that is the random walk centrality of node $i$, where $\tau_{i}^{\alpha}$ is defined as $\displaystyle\sum_{t=0}^{\infty}\{p_{ii}^{\alpha\alpha}(t)-p_i^\alpha\}/p_{i}^{\alpha}$. Combining Eq. \ref{poft}, $\tau_{i}^{\alpha}$ is given by
\begin{equation}
\tau_{i}^{\alpha}=\sum_{k=2}^{NL}\dfrac{1}{1-\lambda_{k}}\dfrac{\phi_{ki}^{2}}{\phi_{1i}^{2}}
\end{equation}

\section*{Acknowledgements}

This work is partially supported by the Chinese National Science Foundation under Grant No.11201017, No.11290141 and No.11401396. Q.G also thanks China Scholarship Council (No.201406020055) for financial support. This work has also been partially supported by a DGA Grant to the group FENOL and by the EC FET-Proactive Project Multiplex (grant 317532).

\section*{Author contributions}

All authors, Q.G, E.C, Z.Z and Y.M, have equally contributed to the conceptualisation of this work, numerical work and its analysis, the creation of all the pictures, and to the writing and revision of the manuscript.

\section*{Additional information}


\textbf{Competing financial interests}
The authors declare no competing financial interests.




\end{document}